\documentclass[12pt,preprint]{aastex}

\slugcomment{To appear in the Astrophysical Journal (Part 1)}

\lefthead{Skinner et al.}
\righthead{EZ CMa}

\begin{document}

\title{{\em XMM-Newton} and {\em VLA} Observations of the Variable \\ 
            Wolf-Rayet Star EZ CMa: Evidence for a Close Companion?   }

\author{Stephen L. Skinner}
\affil{CASA, Univ. of Colorado, Boulder, CO 80309-0389 }

\author{Svetozar A. Zhekov}
\affil{Space Research Institute, Moskovska str. 6,
Sofia-1000, Bulgaria}

\author{Manuel G\"{u}del}
\affil{Paul Scherrer Institute, W\"{u}renlingen and Villigen,
CH-5232 Switzerland}

\and 

\author{Werner Schmutz}
\affil{Physikalisch-Meteorologisches Observatorium Davos,
Dorfstrasse 33, 
CH-7260 Davos Dorf, Switzerland}


%
\newcommand{\ltsimeq}{\raisebox{-0.6ex}{$\,\stackrel{\raisebox{-.2ex}%
{$\textstyle<$}}{\sim}\,$}}
\begin{abstract}
We present new  X-ray and radio observations of the Wolf-Rayet star
EZ CMa (HD 50896) obtained with {\em XMM-Newton} and the {\em VLA}.
This WN4 star exhibits optical and UV variability at a period of
3.765 d whose cause is unknown. Binarity may be responsible but
the existence of a companion has not been proven.
The radio spectral energy distribution of EZ CMa determined from
{\em VLA} observations at five frequencies is in excellent agreement
with predictions for free-free wind emission and the ionized mass-loss
rate allowing for distance uncertainties is 
$\mathrm{\dot{M}}$ = 3.8 ($\pm$2.6) $\times$ 10$^{-5}$ 
M$_{\odot}$ yr$^{-1}$. 
The CCD X-ray spectra show prominent Si XIII and S XV emission
lines and can be acceptably modeled as an absorbed multi-temperature 
optically thin plasma, confirming earlier {\em ASCA} results.
Nonsolar abundances are inferred with Fe notably deficient.
The X-ray emission is dominated by cooler plasma
at a temperature  kT$_{cool}$ $\approx$ 0.6 keV, but a harder
component is also detected and the derived temperature is 
kT$_{hot}$ $\approx$ 3.0 - 4.2 keV if the emission is thermal.
This is too high to be explained by radiative wind shock models and
the X-ray luminosity of the hard component is three orders of magnitude 
lower than expected for accretion onto a neutron star companion.
We show that the hard emission could be produced by the Wolf-Rayet wind
shocking onto a normal (nondegenerate) stellar companion at close
separation. Finally, using comparable data sets we demonstrate that
the X-ray and radio properties of EZ CMa are strikingly similar to
those of the WN5-6 star WR110. This similarity points to common
X-ray and radio emission processes in WN stars and discredits the
idea that EZ CMa is anomalous within its class.
\end{abstract}




%
\newpage

\section{Introduction}
Theories of the origin of X-ray emission in massive stars are now 
being reexamined  in light of new discoveries by the {\em XMM-Newton}
and {\em Chandra} observatories. Traditionally, the X-ray
emission of {\em single} OB and Wolf-Rayet (W-R) stars without companions
has been attributed to shocks distributed throughout their
winds that form as a result of line-driven flow instabilities
(Lucy \& White 1980; Lucy 1982; Baum et al. 1992; Gayley \& Owocki 1995;
Feldmeier et al. 1997; Owocki, Castor, \& Rybicki 1988). Such 
emission is predicted to be relatively soft (kT $<$ 1 keV) and X-ray emission
lines  formed in an optically thick outflowing wind are expected
to be blueshifted and asymmetric due to higher attenuation of
the redward portion of the line by receding material on the far
side of the star (MacFarlane et al. 1991; Owocki \& Cohen 2001).

Recently obtained X-ray grating spectra of several OB  stars reveal
properties that  conflict with the predictions of traditional
radiative wind shock theory. In particular, the emission lines of
the O9 supergiant $\zeta$ Ori (Waldron \& Cassinelli 2000) and the
B0V star $\tau$ Sco (Cohen et al. 2002) are unshifted and show no
obvious asymmetries. The lack of asymmetries is difficult to 
explain if the lines are indeed formed in shocked winds (Owocki \&
Cohen 2001).  Also puzzling is the presence of
narrow high-temperature Fe XXIII and Fe XXIV lines in $\tau$ Sco
indicative of hot plasma well in excess of $\sim$1 keV.
In contrast, much-needed support for the radiative
wind shock paradigm has been received from {\em Chandra}
and {\em XMM-Newton} grating observations of the O4 supergiant
$\zeta$ Puppis, whose emission lines are blueshifted and 
asymmetric (Cassinelli et al. 2001; Kahn et al. 2001). 

Although grating observations are providing stringent tests of
wind shock theories in OB stars, the observational 
picture for  W-R stars is much less complete due to the 
lack of suitable targets bright enough for X-ray grating observations.
At current sensitivity levels, grating spectra of sufficient quality for 
emission line analysis can only be obtained for a few of the
brightest W-R $+$ OB binary systems such as $\gamma^{2}$ Velorum
(Skinner et al. 2001; Dumm et al. 2002). Such binaries provide
crucial information on extrastellar emission from colliding wind
shocks between the stars (Cherepashchuk 1976; Prilutskii \& Usov
1976; Usov 1992), but this emission is difficult to 
decouple from any intrinsic stellar emission that may be
present since the binary components are usually not spatially
resolved in X-rays.

We are thus undertaking an observing program using {\em XMM-Newton}
aimed at acquiring moderate resolution CCD spectra of fainter 
W-R stars that are currently beyond the reach of gratings. 
These include high-interest objects such as the putatively
single WN5-6 star WR 110 (Skinner et al. 2002, hereafter
SZGS02) and possible
binary systems such as EZ CMa discussed here. The large 
effective area of {\em XMM-Newton} is particularly well-suited
for obtaining good-quality CCD spectra of such fainter objects
in reasonable exposure times. Although CCD spectra do not provide
the detailed information on line widths and profiles that can only
be obtained with gratings, they are capable of discerning stronger
emission lines and provide meaningful constraints on X-ray absorption
and the overall distribution of plasma with temperature.
Such information is more than adequate to discriminate between 
the relatively cool emission (kT $<$ 1 keV) that is expected from
filamentary shocks distributed throughout the wind and harder emission
at several keV that is predicted for colliding wind  binaries. 
As such, CCD spectra provide an ideal means for identifying 
candidate colliding wind binaries that may be amenable to study
at higher spectral resolution with next-generation X-ray telescopes.

We present here new {\em XMM-Newton} and  {\em VLA} observations 
of the nitrogen-type W-R star EZ CMa (= HD 50896 = WR6). This WN4
star has been extensively studied at all wavelengths and shows optical 
and ultraviolet variability at a well-documented 3.765 day optical period, 
the origin of which is still not understood. Various explanations have
been proposed including an as yet undetected  companion  
(Firmani et al. 1980; Lamontagne, Moffat \& Lamarre 1986; 
Georgiev et al. 1999) and a rotationally modulated wind  (St.-Louis et 
al. 1995).  The {\em XMM-Newton} observations provide broader energy 
coverage and higher signal-to-noise (S/N) spectra than in previous
observations, giving more accurate measurements  of the X-ray
absorption and plasma temperature distribution for comparison with
emission models. The {\em VLA} data yield the first reliable determination
of the radio spectral index based on single-epoch multifrequency data. \\

\section{Previous X-ray and Radio Observations}

EZ CMa has been observed in X-rays by 
{\em Einstein} (Moffat et al. 1982; White \& Long
1986), {\em ROSAT} (Willis \& Stevens 1996 = WS96),
and {\em ASCA} (Skinner, Itoh, \& Nagase 1997).
The {\em Einstein} and {\em ROSAT} observations
showed day-to-day X-ray variability at levels of
$\leq$30\% but the reality of short-term
($\leq$ 1 hour) variability reported by
White \& Long has been questioned in subsequent
work (WS96). Moffat et al. claimed that weak 
modulation at the 3.76 d optical period was
present in {\em Einstein} data , but
a sequence of eight {\em ROSAT} exposures 
failed to confirm this (WS96).
Low signal-to-noise {\em ASCA} spectra revealed
weak emission lines (Si XIII and S XV) and provided
the first clear evidence that the X-ray emission 
is due to an absorbed multi-temperature 
optically thin plasma. 
The {\em ASCA} observation was not sensitive
enough to detect any photons above $\sim$5 keV,
but the more sensitive {\em XMM-Newton} observation
discussed here shows that such emission is present.

At radio frequencies Hogg (1989) obtained an
average 6 cm flux density S$_{4.89}$ = 1.05 $\pm$ 0.08
mJy from four VLA observations spanning 3.3 years,
with no significant variability. Contreras \&
Rodriquez (2000) analyzed four deep 3.6 cm VLA
observations in both A and D configurations
obtained during a $\approx$6 month interval.
The source was unresolved and flux densities
in the range S$_{8.44}$ =  1.38 $\pm$ 0.03 
to 1.62 $\pm$ 0.03 mJy were obtained. Given
the small flux uncertainties, their results 
could signal episodic variability. Higher
frequency measurements were made by Altenhoff
et al. (1994) at 250 GHz, obtaining S$_{250}$
= 17 $\pm$ 3 mJy. A 230 GHz SEST observation by
Leitherer \& Robert (1991) gave S$_{230}$ =
14.1 $\pm$ 3.1 mJy. \\

\section{Observations}

\subsection{XMM-Newton Observations }
{\em XMM-Newton} observed EZ CMa on 29 - 30 October 2001
as summarized in Table 1.
The X-ray telescope is described by Jansen et al. (2001)
and our analysis focuses on CCD imaging spectroscopy
with the  European Photon Imaging Camera (EPIC). Data
were obtained simultaneously with the EPIC-PN
camera (Str\"{u}der et al. 2001) and two identical EPIC-MOS cameras 
(MOS-1 and MOS-2; Turner et al. 2001). We used full-window mode and
the thick optical blocking filter. The PN and MOS cameras provide
a $\approx$30$'$ diameter field-of-view and energy coverage
from $\approx$0.2 - 15 keV, moderate energy
resolution (E/$\Delta$E $\approx$ 20 - 50), and 
$\approx$5$''$ FWHM angular resolution. Grating spectrometer
data were also acquired but lacked sufficient counts
for analysis.

Data reduction followed standard procedures using the 
{\em XMM-Newton} Science Analysis System software (SAS vers. 5.2).
Pipeline-processed events files generated using the most current
calibration data were filtered with EVSELECT to select good 
event patterns. Spectra and light curves were extracted from
the filtered events lists within circular regions centered on
EZ CMa of radii $\approx$40$''$ (PN) and $\approx$52$''$ (MOS).
The PN extraction radius was limited to $\approx$40$''$ in order
to exclude photons from an adjacent CCD, thus minimizing calibration
uncertainties. The $\approx$52$''$ extraction radius for MOS
excludes a weak field source lying $\approx$57$''$ to the south (Fig. 1).
Background was extracted from source-free regions of the same size 
on the same CCD as the source.
Spectra were rebinned to a minimum of
20 counts per bin for analysis with XSPEC vers. 11.1 (Arnaud 1996)
using  various models. These included  discrete 
temperature optically thin plasma codes (VAPEC and VMEKAL) and 
the differential emission measure (DEM) model C6PVMKL, which is 
an iterative algorithm based on Chebyshev polynomials 
(Lemen et al. 1989). For the hard component, bremsstrahlung
and power-law models
were also examined. All models included an absorption component
based on Morrison \& McCammon (1983) cross sections.

\subsection{VLA Observations}

EZ CMa was observed with the 
NRAO \footnote{The National Radio Astronomy Observatory (NRAO) is 
a facility of the National Science Foundation operated
under cooperative agreement by Associated Universities
Inc.}
VLA during a 3.5 hour interval on 1999 Oct. 19
with the array in hybrid BnA configuration, as summarized in Table 2.
It was observed and detected at five frequencies:
1.42 GHz (21 cm), 4.86 GHz (6 cm), 8.44 GHz (3.6 cm),
14.94 GHz (2 cm), and 22.46 GHz (1.3 cm). 
The observations were made  with the full array (minus two 
inoperable antennas) at each frequency
in scans of $\approx$10 - 12 minutes duration interleaved with
scans of the phase calibrator 0608$-$223. The primary flux 
calibrator 3C286 was  observed at each frequency.

Data were edited and calibrated using the AIPS 
\footnote{Astronomical Image Processing System (AIPS)
is a software package developed by NRAO.}
software package. Maps were produced in both total
intensity (Stokes $I$) and circularly polarized
intensity (Stokes $V$) using the AIPS task IMAGR with 
natural weighting. 
Both peak and total (integrated) fluxes were measured
in cleaned maps using the AIPS tasks TVSTAT (pixel 
summation within a region defined by the  2$\sigma$ contour)
and IMFIT (Gaussian source model). These two methods 
gave good agreement,  differing by no more than the 
RMS noise in the image.

\section{Results }

\subsection{X-ray Properties of EZ CMa}

We summarize below the main observational results 
from analysis of the images, light curves, and 
CCD spectra.

\subsubsection{X-ray Images}

Figure 1 shows the inner region of the   
combined  MOS1 and MOS2 images in broad-band
(0.3 - 10 keV) and hard-band (5 - 10 keV) energy
filters. EZ CMa is clearly detected in both bands
and the X-ray position in Table 3 is offset by only
0.$''$59 from the {\em Hipparcos} optical position
(Perryman et al. 1997).
A weak field source lying 57.4$''$ south of EZ
CMa is visible but there is no SIMBAD counterpart
within 30$''$ of this faint source. 

The hard-band emission is also seen in the PN image, 
which shows 41 net counts (S/N = 5.8) in 
the 5 - 10 keV range. Thus, the hard-band emission is
present in all three detectors and comprises $\approx$1.5\%
of the total counts in each detector. The hard-band 
emission is not extended and the hard emission peak lies
within 0.$''$6 of the EZ CMa {\em Hipparcos} position.
This is the first detection of X-ray photons above 5 keV 
in EZ CMa and is attributable to the larger effective
area of {\em XMM-Newton}.

\subsubsection{X-ray Light Curves}

Figure 2 shows the MOS-1 and MOS-2 light curves
in the 0.3 - 8 keV range. This restricted 
energy range reduces background and thus 
improves sensitivity to any real variability.
The average count rate in each MOS is 
0.12 $\pm$ 0.02 ($\pm$1$\sigma$) counts s$^{-1}$.
There are no MOS fluctuations greater than 
$\pm$2.1$\sigma$ and there is no clear 
correspondence between the largest fluctuations
in one detector and the other.
A fit of an assumed constant 
count rate source to each light curve binned at
512 s gives P(const) =
0.92 ($\chi^2$/dof = 13.4/22) for MOS-1 and 
P(const) = 0.87 ($\chi^2$/dof = 15.5/23) for MOS-2.
Reducing the bin size to 256 s gives nearly
identical values P(const) = 0.92 and 0.86
for MOS-1 and MOS-2. 

By comparison, the PN light curve has an average
count rate of  0.26 $\pm$ 0.03 counts s$^{-1}$.
At a binsize of 256 s there are no fluctuations
greater than $\pm$2.2$\sigma$ and a constant
count rate fit gives P(const) = 0.62 
($\chi^2$/dof = 32.9/36).
Light curves in soft and hard-bands were also
generated and showed similar statistics. For
example, the PN light curve in the 
2.5 - 8.0 keV band gave P(const) = 0.78
($\chi^2$/dof = 4.75/8) when binned at 800 s
intervals ($\approx$20 cts/bin).

In summary, we find no large amplitude
($\geq$3$\sigma$) variability during the 
3.4 hour observation, but lower level
fluctuations at the $\pm$2$\sigma$ level
are present. The probability that these
fluctuations represent real variability
is P(var) = 0.08 - 0.13 in MOS and
P(var) =  0.38 in PN. Thus, real
low-level variability is not totally
ruled out but is considered unlikely.

A comparison of {\em XMM-Newton} and {\em ASCA} 
fluxes does suggest that longer term variability
is present.
The observed (absorbed) flux in the 0.5 - 10 keV range
measured by {\em ASCA} in October 1995 was F$_{x}$(0.5 - 10 keV) =
1.26 $\times$  10$^{-12}$ ergs cm$^{-2}$ s$^{-1}$ (Skinner,
Itoh, \& Nagase 1997). In this
same energy range the flux observed by {\em XMM-Newton}  
is F$_{x}$(0.5 - 10 keV) = 0.96 $\times$ 10$^{-12}$
ergs cm$^{-2}$ s$^{-1}$. 
This represents a 24\% decrease, some of which could be
due to absolute flux calibration differences between
{\em XMM-Newton} and {\em ASCA}. Thus, any real 
variability is within the day-to-day fluctuation range
of $\leq$30\% found in previous {\em ROSAT} monitoring (WS96).

\subsubsection{X-ray Spectra}

The EPIC-PN spectrum is shown in Figure 3. Several
emission lines are visible including 
Mg XI (log T$_{max}$ = 6.8), Si XIII 
(log T$_{max}$ = 7.0),  S XV (log T$_{max}$ = 7.2)
and possible Fe lines in the crowded region
near 1 keV. The Si and S lines confirm earlier
detections in {\em ASCA} spectra (Fig. 3 of Skinner,
Itoh, \& Nagase 1997).

Based on a comparison of several  models fitted to all three
spectra simultaneously, we conclude that the best fit
is obtained with an absorbed two-temperature optically thin plasma
model (2T VAPEC) using nonsolar abundances (Sec. 4.1.4).
Isothermal models are unacceptable.
The best-fit parameters for the
2T VAPEC model are given in Table 3, and an overlay
of this model on the MOS1 spectrum is shown in Figure 4.
The emission measure is dominated by a cool component
at kT$_{cool}$  $\approx$  0.6 keV, but hotter emission is
also present. Assuming that the hotter emission is
thermal, its derived temperature is kT$_{hot}$ = 3.5
[3.0 - 4.2] keV, where brackets enclose the 90\% confidence 
interval. About one-half of the observed (absorbed) flux is
due to the hot component (Table 3). As shown in Figure 5,
essentially all of the flux above 2.5 keV is due to 
the hot component. For comparison, we have also fitted the
{\em ASCA} SIS0 spectrum obtained in October 1995 with the 
2T VAPEC model. Very similar results were obtained and the
best-fit values of N$_{H}$, kT$_{cool}$, and kT$_{hot}$ 
all lie within the 90\% confindence ranges determined
from {\em XMM-Newton} spectra (Table 3). 

Because of the rather high temperature of the hot component
and the low S/N ratio above 4 keV, the
hot component can also be modeled as bremsstrahlung or
a power-law. Replacing the hot optically thin plasma component 
in the 2T VAPEC model with a BREMSS  model in XSPEC gives 
kT$_{brem}$ = 3.4 [3.0 - 4.1] keV and no change in the reduced 
$\chi^2$.  Thus, a bremsstrahlung model for the hard component
is essentially identical to an optically thin plasma model
as expected at these higher temperatures. 
If a power-law model with an energy dependence
E$^{-q}$ is used for the hot
component then the best-fit photon power-law index 
is $q$ = 2.5 and the reduced $\chi^2$ increases by 1.6\%
over the 2T VAPEC value. Issues related to power-law models 
were discussed in SZGS02 and also apply here.

The presence of a cool and hot emission component is 
confirmed in the emission measure distribution derived
from a fit to all three spectra with the C6PVMKL
model, as shown in Figure 6. The cool DEM component
has a maximum contribution at kT$_{cool}$  $\approx$
0.55 keV, in good agreement with 2T models. The DEM
also shows a turnup above $\sim$4 keV signalling
hotter plasma. However, the precise temperature of
the hot component cannot be reliably determined from
DEM models due to low signal-to-noise above 4 keV.

Some soft emission below 0.5 keV appears to be 
present in the PN spectrum (Fig. 3), but is not
as obvious in the MOS spectra (Fig. 4). PN images
in the soft 0.2 - 0.5 keV range show unresolved emission
at the stellar position but there is no indication
of extended emission from the surrounding ring
nebula S 308, in agreement with {\em ROSAT}
findings (WS96). 
The best-fit 2T VAPEC model (Table 3) with a single
absorption component at N$_{H}$ = 4 $\times$ 10$^{21}$ 
cm$^{-2}$ underestimates the flux in the PN spectrum by about
20\% ($-$1.8$\sigma$) in the 0.3 - 0.43 keV range (see 
also Fig. 4 of WS96). However, this model provides a
satisfactory MOS fit at 0.4 keV. The soft residual
in the PN fit can be removed by adding a third optically 
thin plasma component to the  2T VAPEC model at an
uncertain temperature kT$_{soft}$ $<$ 0.3 keV viewed 
under low  ($\approx$interstellar) absorption
N$_{H}$(ISM) $\approx$  5 $\times$ 10$^{20}$ 
cm$^{-2}$. Thus, there could be a contribution
at low energies from very soft emission which
would most likely originate far out in the wind.
But, we are unable to state with confidence that
such very soft emission is present
because of PN calibration uncertainties below 0.5 keV 
and the reduced sensitivity to soft emission that
results from use of the thick optical blocking filter. \\

\subsubsection{Abundances}

We are unable to obtain satisfactory spectral fits
using either solar abundances (Anders \& Grevesse
1989, hereafter AG89) or the  WN abundances given in 
Table 1 of van der Hucht, Cassinelli, \& Williams
1986 (VCW86). An absorbed 2T VAPEC model using solar
abundances gives $\chi^2$/dof = 465.5/223 = 2.09 and
the same model with  WN abundances  gives
$\chi^2$/dof = 383.4/223 = 1.72. Although neither
fit is acceptable, the  WN abundances yield 
somewhat better results and we have thus used them as our 
reference for further variable abundance fitting. 
These canonical WN abundances reflect chemical composition
changes that are thought to occur in WN stars as a result
of advanced nucleosynthesis including hydrogen depletion
and the enhancment of helium and nitrogen (Willis 1996).

The spectral fits are significantly improved by allowing
the abundances of N, Ne, Mg, Si, S, and Fe to vary
relative to the VCW86 values. Also, the fit to the 
weak feature near 3.15 keV can be improved by adding
Ar to the canonical WN abundance table and letting
it vary. This weak feature could be Ar XVII, but the 
S XVI line at 3.11 keV may also be 
contributing.  A comparison of 2T VAPEC
and C6PVMKL models suggests that  Fe is underabundant
in EZ CMa by about a factor of four relative to the
values in VCW86 (even when Ne is also allowed to vary)
and that S is overabundant. Specifically,
2T VAPEC fits give Fe = 0.23 [0.15 - 0.32] and
S = 2.0 [1.4 - 2.9] relative to the VCW86 values,
which are
by number: Fe/H = 1.904 $\times$ 10$^{-3}$ and
S/H = 7.600  $\times$ 10$^{-4}$. Similar values are
obtained with a 2T VMEKAL optically thin plasma model.
If cosmic abundances (AG89) are
used as a reference  
then the underabundance of Fe persists.

\subsubsection{Absorption and A$_{v}$}

Acceptable fits give an equivalent neutral hydrogen column density
N$_{H}$ = 4.0 [3.4 - 4.4] $\times$ 10$^{21}$  cm$^{-2}$.
This value  corresponds to a visual extinction A$_{v}$ = 1.8 [1.5 - 2]
using the conversion of Gorenstein (1975), which is at least
twice as large as the values A$_{v}$ = 0.00 - 0.82 determined
from optical studies (van der Hucht 2001, hereafter vdH01).
However, if we adopt an absolute visual magnitude M$_{v}$ = 
$-$3.5 as typical for WN4 stars (Fig. 4 of vdH01), $v$ = 6.94
(vdh01) and the {\em Hipparcos} distance of 575 pc, then 
we obtain A$_{v}$ = 1.64, which is in good agreement with 
the value obtained above from the X-ray spectra. Thus, 
the larger A$_{v}$ determined from X-ray spectra is not
unrealistic but two factors should be kept in mind.
First, a typical value of M$_{v}$ is difficult to define
for WN stars since there is no definite relation between
spectral type and bolometric luminosity.  
Second, the {\em Hipparcos} 
distance is quite uncertain as evident from the parallax
measurement $\pi$ = 1.74 $\pm$ 0.76 mas (Perryman et al. 
1997).

\subsection{Radio Properties}

The VLA position (Table 2) is in excellent agreement  
with the Hipparcos position, with a 
radio $-$ optical offset  $\Delta$RA = $+$0.001 s and 
$\Delta$DEC = $+$0.07$''$. The source is unresolved down
to the smallest synthesized beamsize of 0.$''$4 $\times$
0.$''$3 as indicated by the good agreement between the peak
and total fluxes.  
The total flux at 4.86 GHz is nearly 
identical to the value S$_{4.86}$ = 1.05 $\pm$ 0.08 mJy 
measured in 1980 - 1983 with the VLA (Hogg 1989).
No circular polarization was
detected and the most stringent upper limit on the fractional
circular polarization ($\pi_{c}$)  from the 8.44 GHz Stokes V
image is $\pi_{c}$ $\leq$ 0.067 (3$\sigma$).

As shown in Figure 6, the total fluxes are
well-approximated by a power-law of the form S$_{\nu}$ 
$\propto$ $\nu^{\alpha}$, where $\alpha$ = $+$0.69$\pm$0.05 (90\%
confidence errors). This spectral index agrees very well 
with that expected for free-free emission
from  a spherical ionized constant-velocity wind (Wright \& Barlow 1975).
It is also consistent with the value $\alpha$ = $+$0.64$\pm$0.06
derived by Leitherer \& Robert (1991) using non-contemporaneous
flux measurements.
Extrapolating the total 22.4 GHz flux to 250 GHz using 
$\alpha$ = $+$0.69 gives a predicted value S$_{250}^{(pred)}$ = 16.6 $\pm$
0.8 mJy, which is consistent  with the measurement 
S$_{250}$ = 17 $\pm$ 3 mJy 
obtained by Altenhoff, Thum, \& Wendker (1994). Thus, there is
good reason to believe that the $\alpha$ = $+$0.69 power-law
extends to frequencies an order of magnitude higher than 
observed here with the VLA.

\subsubsection{Mass Loss Rate}
The ionized mass loss rate for an assumed constant-velocity
wind can be estimated using the
result of  Wright \& Barlow (1975), namely 
$\mathrm{\dot{M}}$ = 
C$_{0}$v$_{\infty}$S$_{\nu}^{0.75}$d$_{kpc}^{1.5}$ M$_{\odot}$ 
yr$^{-1}$,
where C$_{0}$ = 0.095$\mu$/[Z$\sqrt{\gamma g \nu}$].
Here, v$_{\infty}$ (km s$^{-1}$)  is the terminal 
wind speed, S$_{\nu}$ (Jy) is the observed radio flux at
frequency $\nu$ (Hz), d$_{kpc}$ (kpc) is the stellar distance,
$\mu$ is the mean atomic weight per nucleon, Z is the
rms ionic charge, $\gamma$ is the mean number of free 
electrons per nucleon, and g is the free-free Gaunt factor.
To evaluate this expression we use the highest 
signal-to-noise radio detection at  $\nu$ = 8.44 GHz,
S$_{8.44}$ = 1.74 mJy (Table 2), v$_{\infty}$ = 1700
km s$^{-1}$ (Hillier 1987; Prinja, Barlow, and Howarth 1990), Z = 1,
$\gamma$ = 1, and g = 4.78
at 8.44 GHz from the approximation for the free-free Gaunt factor
given in Eq. [8] of Abbott et al. (1986), assuming a temperature
at the radio photosphere of T = 10000 K (VCW86). These values give 
$\mathrm{\dot{M}}$ = 
6.85 $\times$ 10$^{-6}$ $\mu$ d$_{kpc}^{1.5}$ M$_{\odot}$ yr$^{-1}$.
Assuming $\mu$ = 3.9 for a WN star (VCW86) this becomes
$\mathrm{\dot{M}}$ = 2.67 $\times$ 10$^{-5}$d$_{kpc}^{1.5}$
M$_{\odot}$ yr$^{-1}$. The wind luminosity is
L$_{wind}$ = (1/2)$\mathrm{\dot{M}}$v$_{\infty}^{2}$ =
2.4 $\times$ 10$^{37}$d$_{kpc}^{1.5}$ ergs s$^{-1}$.

Using the range of distance estimates from 
the Hipparcos value of 0.575 kpc
to  1.8 kpc from Howarth  \& Schmutz (1995) to evaluate
the above quantities we obtain
$\mathrm{\dot{M}}$ = (1.2 - 6.4) $\times$ 10$^{-5}$
M$_{\odot}$ yr$^{-1}$ and L$_{wind}$ = 
(1.1 - 5.9) $\times$ 10$^{37}$ ergs s$^{-1}$.
Smaller values of $\mathrm{\dot{M}}$ are obtained using
values of Z $>$ 1 and $\gamma$ $>$ 1 obtained from
specific atmospheric models (VCW86). However, the
values of Z and  $\gamma$ are quite sensitive to poorly-known stellar
parameters and we have thus adopted Z = 1 and $\gamma$ = 1 in
the above calculation.

\section{Discussion}

The new observational results discussed above provide the
most detailed picture to date of the X-ray and centimeter
radio properties of EZ CMa. Below, we make comparisons
with previous studies and comment on specific
emission models.

\subsection{WN Stars: Comparative Spectroscopy}

The {\em XMM-Newton} and VLA data  for
EZ CMa provide a good basis for comparison with similar
data recently obtained for WR 110 (SZGS02). For the
first time we are able to directly compare the X-ray
and radio properties of two WN stars having similar
spectral types using analogous data sets. 

Table 4 summarizes the X-ray and radio properties
of EZ CMa and WR 110, and their X-ray spectra are
compared in Figure 8. Overall, the two stars are
strikingly similar and the same physical processes
are very likely responsible for the X-rays and
radio emission in both stars. One notable difference 
is the larger N$_{H}$ for WR 110, attributable to its 
stronger interstellar absorption toward the Galactic
center (SZGS02).  The uncertain distance for
EZ CMa introduces some ambiguity into the calculation
of $\mathrm{\dot{M}}$ and L$_{x}$. But, as Table 4 shows,
if an intermediate distance of 1.2 kpc is adopted for
EZ CMa  then $\mathrm{\dot{M}}$ and L$_{x}$ agree with
WR 110 to better than a factor of two.

\subsection{Radiative Wind Shocks}

The {\em XMM-Newton} spectra provide some constraints on
wind shock models. The dominant cool emission component
peaking near kT$_{cool}$ $\approx$ 0.6 keV can potentially
be explained by radiative wind shock models, but the 
hot component requires a different explanation.

Assuming that the cool component is due to shocks distributed
throughout the wind, then the observed X-ray temperature 
provides constraints on the shock speed v$_{s}$ using 
the adiabatic shock formula 
kT$_{s}$ = (3/16)\={m}v$_{s}^2$. For a helium-rich
WN wind \={m} = (4/3)m$_{p}$ where m$_{p}$ is the
proton mass. The value kT$_{cool}$ $\approx$ 0.6 keV
gives a typical shock speed v$_{s}$ $\approx$ 480
km s$^{-1}$ but the range in temperatures of 
$\approx$0.3 - 0.8 keV inferred from the FWHM of
the DEM model implies a range of shock speeds 
v$_{s}$ $\approx$ 340 - 550 km s$^{-1}$. These 
values overlap the range of shock velocity jumps
$\Delta$v = 500 - 1000 km s$^{-1}$ derived in
numerical simulations of radiatively driven winds
(Owocki et al. 1988).

The average filling factor $f$ of X-ray emitting plasma in the 
wind can be estimated using the procedure given in SZGS02 and
the mass-loss parameters from Section 4.2.1. We define 
$f$ = EM$_{x}$/EM$_{tot}$ where EM$_{x}$ is the volume emission measure 
of the X-ray emitting plasma and EM$_{tot}$ is the total volume emission
measure in the wind. For a He-dominated wind we obtain
$f$ = 5.34 $\times$ 10$^{-3}$(R$_{*}$/R$_{\odot}$)$norm$/d$_{kpc}$,
where $norm$ is the normalization factor from XSPEC VAPEC models.
Since only the cool X-ray component can be attributed to 
radiative wind shocks, we set $norm$ = $norm_{cool}$ from
Table 3 and obtain
$f$ = 1.8 $\times$ 10$^{-7}$(R$_{*}$/R$_{\odot}$)/d$_{kpc}$.
For radii of a few solar radii (Hillier 1987) and the
range of estimated distances 0.58 - 1.8 kpc, the filling factor need
not be larger than $f$ $\sim$ 10$^{-6}$.

The emergent cool emission detected
by {\em XMM-Newton} must originate at large distances from the
star unless the wind is clumped. Using the same procedure as
in SZGS02 along with an intermediate distance d = 1.2 kpc and
the mass loss parameters in Sec. 4.2.1, the radius of optical
depth unity at 1 keV is R$_{\tau = 1}$(E = 1 keV) = 
1.58 $\times$ 10$^{14}$ cm $\approx$ 10.6 AU. Assuming 
R$_{*}$ $\approx$ 2  R$_{\odot}$ as a representative
value (Hillier 1987; Hamann \& Koesterke 1998), then
R$_{\tau = 1}$(E = 1 keV) $\approx$ 1130 R$_{*}$.
The harder emission could be coming from much smaller
radii. Specifically, R$_{\tau = 1}$(E = 4 keV) 
$\approx$ 0.3 AU assuming that the wind absorption
cross-section for X-rays $\sigma_{w}$ scales 
with energy according to $\sigma_{w}$ $\propto$
E$^{-2.5}$ (Fig. 1 of Ignace, Oskinova, \& Foullon
2000).  

The shock speeds, filling factor and value of R$_{\tau = 1}$(E = 1 keV)
computed above are nearly identical to those derived previously for 
WR 110 (SZGS02). The unit optical depth calculations for EZ CMa and
WR 110 suggest that soft X-rays ($\sim$1 keV) emerge at many hundreds
of stellar radii in WR stars assuming homogeneous winds, but smaller
emergent radii would be possible for clumped winds. Similar conclusions
have been reached for some O-type stars (e.g. Hillier et al. 1993).
Detailed hydrodynamic simulations are now needed to determine if
instability-generated wind shocks can persist to hundreds of radii
in WR stars, analogous to those recently undertaken for OB stars
by Runacres \& Owocki (2002).

\subsection{On the Possibility of a Compact Companion}

Binarity has been suggested as one possible means of
explaining the 3.765 day optical variability of EZ CMa.
An estimate of the mass of the putative companion
M$_{comp}$ = 1.3 ($\pm$0.4) M$_{\odot}$ derived by
Firmani et al. (1980) raised speculation that 
EZ CMa could have a compact companion (c), making it
a rare WR $+$ c system (Lamontagne et 
al. 1986; White \& Long1986). But, it was shown
that the X-ray luminosity of EZ CMa is about three
orders of magnitude lower than the accretion luminosity
L$_{x}^{(acc)}$ $\sim$10 $^{36}$ ergs s$^{-1}$ expected
for accretion of the W-R wind onto a neutron star
(Stevens \& Willis 1998). Similar arguments against
a black hole companion based on {\em ASCA} luminosities
were given by Skinner et al. (1997).

The {\em XMM-Newton} results  confirm 
the above luminosity deficit, giving an unabsorbed luminosity 
L$_{x}$ = (0.2 - 10 keV) = 3.46 $\times$
10$^{32}$d$_{kpc}^{2}$ ergs s$^{-1}$. At the upper end of
current distance estimates, d = 1.8 kpc and 
L$_{x}$ = 10$^{33.0}$ ergs s$^{-1}$. If only the 
contribution of the hard component is considered, then
L$_{x,hard}$(0.2 - 10 keV) = 10$^{32.5}$ ergs s$^{-1}$.
Thus, if a
neutron star companion is present then some mechanism
such as rapid rotation near breakup (Davidson \& Ostriker 1973) 
is needed to inhibit accretion.

The strong similarity between EZ CMa and WR 110 (Table 4) 
presents a new challenge for the compact companion
hypothesis.  WR 110 has so far shown no clear signs of
binarity or periodic optical variability and has not
previously been proposed as a candidate WR $+$ c system. 
If EZ CMa has a compact companion, then it is not clear
why its X-ray and radio properties would so closely
mimic those of another WN star for which evidence of
a compact companion is lacking.

\subsection{On the Possibility of a Normal Stellar Companion}

An interpretation of the X-ray emission in terms of accretion
onto a compact companion is questionable on the above grounds.
However, it is  more difficult to rule out a normal 
(nondegenerate) stellar companion. Such a companion is expected
to be much less massive than the W-R star given that the radial
velocity variations reported by Firmani et al. (1980) are 
of low amplitude.  Arguments for binarity have recently been
strengthened on the basis of long-term  coherent optical
variability seen in data sets spanning
more than 15 years  with a well-determined period 
P = 3.765 $\pm$ 0.0001 days (Georgiev et al. 1999).
Georgiev et al. have argued that if the optical variability
is due to a  stellar companion then it is most likely  
orbiting very close to EZ CMa near the base of the wind
in order to explain variations in the N V lines. 
We show below that a close companion could account for
the hard X-ray emission detected by {\em XMM-Newton}.

To constrain the separation, we assume that P = 3.765 d 
is an orbital period and that the companion mass is much
less than that of the W-R star M$_{wr}$ $\approx$ 
16 M$_{\odot}$  (Hamann \& Koesterke 1998).
Kepler's third law then gives the separation in
AU as a$_{AU}$ $\approx$ 0.12, or equivalently
a $\approx$ 25 R$_{\odot}$. A nearly identical
separation is obtained if one uses the values of
M$_{wr}$ $\approx$  10 M$_{\odot}$ and the 
companion mass M$_{comp}$ $\approx$ 1.3 M$_{\odot}$
adopted by Firmani et al. (1980).

We now assume that
the hard X-ray component is produced by the W-R
wind shocking onto the lower mass companion and
that the W-R wind is dominant. In this case
the contact surface is the surface of the companion
star, as discussed in more detail by Luo, McCray, \& 
MacLow (1990).
Because of the close separation, radiative cooling
may be important (eq. [8] of Stevens et al. 1992; 
eq. [52] of Usov 1992). We thus consider both
adiabatic and radiative shocks.
The close separation also raises
the question of whether hard X-rays could escape
from the overlying W-R wind and be detected.
As already noted (Sec. 5.2), the radius of
optical depth unity at 4 keV is  
R$_{\tau = 1}$(4 keV) $\approx$ 0.3 AU,
so some absorption of the hard X-rays could occur.
However, the absorption will depend critically
on wind properties such as the clumping factor,
and the actual value of R$_{\tau = 1}$ in a
clumped wind would be less than that given
above, which is based on the assumption of a
spherical, homogeneous wind. Thus, without
more specific information on wind geometry and
homogeneity the escape of hard X-rays ($\sim$4 keV)
from radii smaller than 0.3 AU is not precluded.

The inferred radius
of the companion R$_{comp}$ is obtained by equating the 
unabsorbed luminosity of the hard component (Table 3) with the 
predicted shock luminosity, where the predicted 
shock luminosity is different in the adiabatic and
radiative cases (eqs. [79], [80] of Usov 1992, respectively). 
Using the mass loss parameters in Sec. 4.2.1 and the unabsorbed
hard-component flux F$_{x}^{(hot)}$(0.2 - 10 keV) = 7.8 
$\times$ 10$^{-13}$ ergs cm$^{-2}$ s$^{-1}$, the adiabatic
case gives 
a$_{AU}$ = 0.20d$_{kpc}^{0.25}$(R$_{comp}$/R$_{\odot}$)$^{0.75}$.
For a$_{AU}$ = 0.12 and the probable distance range  
d$_{kpc}$ = 0.575 - 1.8 the companion radius is 
R$_{comp}$ $\approx$ 0.4 - 0.6 R$_{\odot}$. A similar
calculation in the radiative case gives 
R$_{comp}$ $\approx$ 0.12 - 0.16 R$_{\odot}$.
In the above, we have used the luminosity
in a specific energy range (0.2 - 10 keV) as an
approximation for the total X-ray luminosity 
integrated over all frequencies. In general, the
luminosity within a specific bandpass will
be less than the value over all frequencies so
the derived values of R$_{comp}$  are in
fact lower limits.

The observed 
temperature kT$_{hot}$ $\approx$ 3.5 [3.0 - 4.2] keV
is also compatible with a shocked companion interpretation.
The observed temperature derived from spectral fits
is an average that includes contributions from the 
hottest shocked plasma at kT$_{s,max}$ along the 
line-of-centers and cooler plasma downstream. Thus,
the observed value will in general be {\em less} than
the maximum temperature. The difference between
the observed temperature and kT$_{s,max}$ will depend
on several poorly known factors including wind chemical 
composition and the importance of radiative cooling.
As a rough estimate, one expects an observed temperature
kT $\approx$ 0.8kT$_{s,max}$ in the adiabatic case
(eq. [83] ff. of Usov 1992). 

For EZ CMa, the maximum predicted temperature for
an adiabatic shock in a He-dominated wind is
kT$_{s,max}$ $\approx$ 7.5 keV (Sec. 5.2), 
assuming that the wind has reached terminal
speed v$_{\infty}$ = 1700 km s$^{-1}$ at the
shock. However, at the close separation of
interest here (a $\approx$ 25 R$_{\odot}$),
the wind may not have reached terminal speed.
Using the velocity profile and stellar radius
R$_{wr}$ = 2.5 R$_{\odot}$ given by Hillier
(1987), we obtain a/R$_{wr}$ $\approx$ 10 and the 
wind is at or near terminal speed when it impacts
the companion. But, using the radius
R$_{wr}$ = 3.5 R$_{\odot}$ and $\beta$ = 3
wind velocity law of Schmutz (1997) then
a/R$_{wr}$ $\approx$ 7.1 and the wind speed is 
v$_{wr}$ $\approx$ 0.64 v$_{\infty}$. In this
case, kT$_{s,max}$ $\approx$ 3.1 keV. 
Thus, based on the rough approximation that the
observed temperature should be $\approx$0.8 kT$_{s,max}$,
we would expect observed values in the range
$\approx$2.5 - 6 keV. The values measured from
spectral fits are well within this range.

We thus conclude that the X-ray luminosity and
observed temperature of the hard component are compatible  
with the shocked companion hypothesis. If the companion
mass is much less than that of the W-R star then
the inferred companion radius is at least 
R$_{comp}$ $\approx$ 0.2 - 0.6 R$_{\odot}$.
This would correspond to a main-sequence (MS) 
M-type star, but clearly such a low mass 
star could not yet have reached the MS if it
formed contemporaneously with the W-R star.
Thus, a pre-main-sequence (PMS) companion would seem
more likely. It may be relevant here that 
some W-R stars such as $\gamma^{2}$ Vel are 
now believed to be associated with PMS objects
(Pozzo et al. 2000; Skinner et al. 2001).
If the putative companion is indeed a late-type
star then interaction of the W-R wind with 
the companion's magnetic field would be an
important factor to include in more detailed
models. 
Our analysis shows that the observed temperature
of the hot component is somewhat less than that
expected for an adiabatic shock if the wind 
has reached terminal speed. This could be an
indication that radiative cooling is important
and the adiabatic approximation is breaking
down at close separation, or perhaps that
the W-R wind has not reached terminal speed
at the shock. Radiative braking of the WR wind
by the the companion star
can also lead to lower X-ray temperatures than
predicted by simple models (Gayley, Owocki, 
\& Cranmer 1997) but this effect would be of
little importance for a lower mass companion
that is less luminous than the WR star.

\section{Summary}

Overall, the X-ray properties derived for EZ CMa using new
{\em XMM-Newton} data are in good agreement with previous
{\em ASCA} results (Skinner et al. 1997). The improved 
signal-to-noise in {\em XMM-Newton} spectra  yields more 
accurate measurements of the X-ray absorption and plasma 
temperatures. The temperature of the dominant cool component
is identical to that obtained with {\em ASCA}, but the hot
component temperature is revised upward by $\approx$35\%
to kT$_{hot}$ $\approx$ 3.5 [3.0 - 4.2] keV.

The origin of the hotter plasma remains
obscure, but neutron star accretion models are difficult
to justify based on X-ray luminosity considerations and
the close similarity between the X-ray and radio properties
of  EZ CMa and WR 110, where the latter is {\em not} a candidate
WR $+$ c system. We have argued that the hard X-rays could 
from the W-R wind shocking onto a normal low-mass stellar
companion, but direct evidence for such a companion
is still lacking. Also, the failure to detect periodic X-ray variability
with {\em ROSAT} is somewhat difficult to reconcile  
with a companion interpretation unless the orbital elements are not
conducive to X-ray variability, as for example in a
low eccentricity orbit. Further broad-band
searches for modulated X-ray
emission might be worthwhile since {\em ROSAT} was not sensitive
to emission above 2.5 keV. The {\em XMM-Newton} spectra show that
about one-third of the observed flux emerges above 2.5 keV
(Table 3) and essentially all of this is due to the hard component.

Until conclusive evidence for binarity is found, alternative
models that are capable of producing hard X-rays in the absence
of a companion should be explored. Of particular interest
are models that incorporate wind - magnetic field interactions
since it has been proposed that the variability of EZ CMa
may be due to corotating structures at or near the surface
(St.-Louis et al. 1995). Possible evidence for magnetic 
activity in the form of short-duration ($\sim$1 hr) optical
variability has also been reported (Duijsens et al. 1996).

The {\em XMM-Newton} spectra also reveal two interesting 
anomalies that deserve further study. First, the absorption
column density derived from {\em XMM-Newton} spectra (Table 3)
is comparable to {\em Einstein} and {\em ASCA} values and
leads to an estimate of A$_{v}$ = 1.8 [1.5 - 2.0] that is at least 
twice as large as previous optical/UV determinations. This suggests
that excess X-ray absorption could be present that is not seen
in the optical/UV, possibly due to the surrounding ring nebula,
or perhaps that A$_{v}$ is slightly underestimated from optical/UV 
studies.  Second,
the {\em XMM-Newton} spectra indicate  that WN abundances may
be substantially different than assumed in earlier work.
In particular, all spectral models examined here give a best-fit
Fe abundance that is a factor of $\sim$4 below the reference
value for WN stars given in VCW86, and a similar depletion 
is noted when using cosmic abundances (AG89) as a reference.
Some caution is advised since the 
low Fe abundance was inferred from moderate resolution
CCD spectra where line blending is present, but the repeatibility
of this result across several different types of models does suggest
that the underabundance is real.

Finally, the close similarities between EZ CMa and WR 110
apparent in Table 4 and Figure 8 represent a first step toward a unified
observational picture of X-ray and radio emission in WN stars.
These close similarities provide strong evidence that the physical 
mechanisms responsible for the  X-ray
and radio continuum emission in these two WN stars are 
identical, even though their optical properties are 
obviously different. If an unseen companion
is responsible for the optical variability and hard X-ray 
emission in EZ CMa then the similarities in Table 4
implicate an unseen companion in WR 110 as well. The
properties of such a companion were derived in our
previous analysis of WR 110 (SZGS02) but are not as
tightly constrained as for EZ CMa since WR 110 has no
known regular optical variability on which to base an
estimate of the orbital period.  
Other W-R stars
which have so far shown no direct evidence for binarity
but which could also harbor unseen companions are those 
with prominent X-ray emission such as the WN4 star
WR1 (HD 4004).


\acknowledgments

This work was supported by NASA grant NAG5-10325. 
Research at PSI has been supported by the Swiss
National Science Foundation under grant
2100-049343.
This work was based on observations obtained with 
XMM-Newton, an ESA science mission with instruments and
contributions directly funded by ESA member states
and the USA (NASA). We thank members of the XMM-Newton,
VLA, and HEASARC (NASA/GSFC) support teams for their
assistance. This research has made use of the SIMBAD
astronomical database operated by CDS at Strasbourg,
France. 

\clearpage
\begin{deluxetable}{ll}
\tablewidth{0pc}
\tablecaption{XMM-Newton Observations of EZ CMa\tablenotemark{a} }
\tablehead{
\colhead{Parameter}      &
\colhead{Value  }     
}
\startdata
Start (UT)                             & 29 Oct 2001 21:35                                    \nl
Stop  (UT)                             & 30 Oct 2001 00:57                                    \nl
Exposure (ksec)                        & 9.5 (PN), 12.1 (per MOS)                             \nl
Count Rate $\pm$1$\sigma$ (c s$^{-1}$) & 0.26 $\pm$ 0.03 (PN),  0.12 $\pm$ 0.02 (per MOS)     \nl
Flux  (10$^{-12}$  ergs cm$^{-2}$ s$^{-1}$)        & 0.97 (2.90)             \nl
\tablenotetext{a} {Count rate and standard deviation are from light curves
binned at 256 s intervals.
Flux is the observed  value in the 0.2 - 10 keV range 
followed in parentheses by the unabsorbed value.}
\enddata
\end{deluxetable}
\clearpage
%
\begin{deluxetable}{llcccc}
\tablewidth{0pc}
\tablecaption{VLA Observations of EZ CMa\tablenotemark{a} }
\tablehead{
\colhead{Frequency} &
\colhead{Beam FWHM } &
\colhead{Duration} &
\colhead{RMS Noise } &
\colhead{Peak Flux} &
\colhead{Total Flux} \\
\colhead{(GHz)         } &
\colhead{(arcsec)} &
\colhead{(min.)} &
\colhead{($\mu$Jy/beam)} &
\colhead{(mJy/beam)} &
\colhead{(mJy)}  \\
}
\startdata
1.42  & 6.1 $\times$ 4.2 & 23 & 56\tablenotemark{b}  &  0.56 & 0.57  \nl
4.86  & 1.7 $\times$ 1.1 & 20 & 39  &  1.01 & 1.02  \nl
8.44  & 0.9 $\times$ 0.7 & 20 & 29  &  1.60 & 1.74  \nl
14.94 & 0.5 $\times$ 0.4 & 31 & 99  &  2.31 & 2.38  \nl
22.46 & 0.4 $\times$ 0.3 & 30 & 150 &  3.16 & 3.15  \nl
\tablenotetext{a}{All data were obtained in BnA configuration
on 1999 Oct 19 from 
1024 - 1341 UT. Observations were obtained at each frequency in
two orthogonal polarization channels, each with a bandwidth of 50 MHz.
Fluxes and beam sizes are from cleaned Stokes I maps. Primary
flux calibrator was 3C286. The radio position of EZ CMa measured
from 8.44 GHz maps is RA(2000) = 06~h 54~m 13.045~s, DEC(2000) =
$-$23$^{\circ}$ 55$'$ 41.94$''$.}
\tablenotetext{b}{The RMS noise at 1.42 GHz is larger than predicted
by sensitivity calculations due to the presence of other bright sources in
the field.}
\enddata
\end{deluxetable}
\clearpage
%
\begin{deluxetable}{ll}
\tablewidth{0pc}
\tablecaption{X-ray Spectral Properties of EZ CMa\tablenotemark{a} \label{tbl-1}}
\tablehead{
\colhead{Parameter}      &
\colhead{Value  }       
}
\startdata
N$_{H}$ (10$^{21}$ cm$^{-2}$)             & 4.0 [3.4 - 4.4]                   \nl
kT$_{cool}$ (keV)                         & 0.59 [0.55 - 0.63]                \nl
kT$_{hot}$  (keV)                         & 3.5 [3.0 - 4.2]                   \nl
norm$_{cool}$                             & 3.41 $\times$ 10$^{-5}$           \nl
norm$_{hot}$                              & 1.04 $\times$ 10$^{-5}$           \nl
EM$_{cool}$/EM$_{hot}$                    & 3.3                               \nl
Abundances                                & varied                            \nl
$\chi^2$/dof                              & 234.5/217                         \nl
F$_{x}$(0.2 - 10 keV)\tablenotemark{b}    & 0.97 (2.90)                       \nl
F$_{x}$(2.5 - 10 keV)\tablenotemark{b}    & 0.31 (0.33)                       \nl
F$_{x}^{(hot)}$(0.2 - 10 keV)\tablenotemark{b}        & 0.49 (0.78)                       \nl
\tablenotetext{a} {X-ray properties are from simultaneous fits of 
PN, MOS1, and MOS2 spectra using an absorbed 2T optically
thin plasma model (2T VAPEC). Brackets enclose  90\% confidence intervals.
Abundances were initialized to the 
WN values given in Table 1  of VCW86 (renormalized
to H = 1.0 in XSPEC) and N, Ne, Mg, Si, S, Fe  abundances were  
then varied to obtain the best fit.
Iron converges to an abundance Fe = 0.23 [0.15 - 0.32]. 
The emission measure
for a He-dominated plasma is EM = 1.9 $\times$ 10$^{59}$d$_{kpc}^{2}$$norm$
cm$^{-3}$. The X-ray position obtained by averaging the
results of all three EPIC cameras is RA(2000) = 06$^{h}$  54$^{m}$ 13.07$^{s}$,
DEC(2000) = $-$23$^{\circ}$ 55$'$ 41.5$''$.} 
\tablenotetext{b} {Fluxes are the observed (absorbed) values followed 
in parentheses by the unabsorbed values in units of 10$^{-12}$ 
ergs cm$^{-2}$ s$^{-1}$.}
\enddata
\end{deluxetable}
\clearpage
%
\begin{deluxetable}{lll}
\tablewidth{0pc}
\tablecaption{Comparison of EZ CMa (WR 6) and WR 110 \label{tbl-1}}
\tablehead{
\colhead{Parameter}      &
\colhead{EZ CMa  }        &
\colhead{WR 110  }
}
\startdata
Spectral type\tablenotemark{a}            & WN4                 & WN5-6       \nl
Distance (kpc)\tablenotemark{a}           & 0.58 - 1.8          & 1.28     \nl
Radio spectral index ($\alpha$)           & $+$0.69 $\pm$ 0.05  & $+$0.64 $\pm$ 0.10   \nl
$\mathrm{\dot{M}}$ (10$^{-5}$ M$_{\odot}$ yr$^{-1}$) & 3.5 (1.2 - 6.5)\tablenotemark{b}  & 4.9         \nl
N$_{H}$ (10$^{22}$ cm$^{-2}$)             & 0.4 $\pm$ 0.05      & 1.05 $\pm$ 0.16        \nl
kT$_{cool}$ (keV)                         & 0.59 $\pm$ 0.04     & 0.55 $\pm$ 0.07       \nl
kT$_{hot}$ (keV)                          & 3.5  $\pm$ 0.6      & $\geq$3     \nl
L$_{x}$ (10$^{32}$ ergs s$^{-1}$)         & 5.0 (1.1 - 11.2)\tablenotemark{b}   & 4.4   \nl
L$_{x}$/L$_{wind}$                        & $\sim$10$^{-5}$     & $\sim$10$^{-5}$  \nl
L$_{6cm}$/L$_{x}$ (10$^{-15}$ Hz$^{-1}$)  & 3.6                 & 5.2  \nl 
\tablenotetext{a} {Spectral types and the distance for WR 110 are from vdH01.
The smaller distance for EZ CMa is from {\em Hipparcos} and the larger is from
Howarth \& Schmutz 1995. The X-ray properties of WR 110 are from SZGS02
except that L$_{x}$ is quoted here in the 0.2 - 10 keV range.}
\tablenotetext{b} {The quoted value is for an intermediate distance of 1.2 kpc, followed
in parentheses by the range of values corresponding to d = 0.58 - 1.8 kpc. 
L$_{x}$ is the unabsorbed  luminosity in the 0.2 - 10 keV range.}
\enddata
\end{deluxetable}
\clearpage

\clearpage

\figcaption{Smoothed  X-ray contour images of the region near 
EZ CMa (arrow) obtained by combining data from the MOS1 and MOS2
detectors. The combined exposure time is $\approx$24.2 ksec and
coordinate overlay is J2000. The
respective energy bands, total EZ CMa source counts (MOS1 $+$ MOS2) 
after background
subtraction, and S/N ratio in each image are:~
{\em left:} broad-band (0.3 - 10 keV), 2933 counts, S/N = 52.1;~ 
{\em right:} hard-band (5 - 10 keV), 44 counts, S/N = 6.0. 
\label{fig1}}

\figcaption{Background-subtracted 
EPIC-MOS light curves of EZ CMa in the 0.3 - 8.0 keV range.
Error bars are 1$\sigma$ and the binsize is 512 s. 
The average count rate is 0.12 $\pm$ 0.02 c s$^{-1}$
per MOS.
\label{fig2} }

\figcaption{Histogram plot of the background-subtracted 
EPIC-PN spectrum of EZ CMa
rebinned to a minimum of 15 counts per bin for display.
\label{fig3} }

\figcaption{Best fit 2T VAPEC model using the parameters
in Table 3 overlaid on the MOS1 spectrum. The spectrum is
rebinned to a minimum of 20 counts
per bin. Residuals are in the sense of data - model.
\label{fig4} }

\figcaption{Contributions of the cool and hot components
to the best-fit 2T VAPEC model of the MOS-1 spectrum.
The hot component dominates the emission above 
$\approx$2.5 keV.
\label{fig5} }

\figcaption{Differential emission measure (DEM) model of EZ CMa
based on a simultaneous fit of the  EPIC
spectra from all three detectors using
an absorbed Chebyshev polynomial model (C6PVMKL). The best-fit column
density is  N$_{H}$ = 4 $\times$ 10$^{21}$ cm$^{-2}$ and abundances
were allowed to vary relative to the canonical WN abundances given
by VCW86.
\label{fig6} }

\figcaption{The radio spectral energy distribution of EZ CMa
based on total fluxes given in Table 1. Error bars are 
the formal errors in total flux computed by the Gaussian
fit AIPS task IMFIT. The solid
line is a best-fit power-law model, which gives a spectral
index $\alpha$ = $+$0.69 $\pm$ 0.05 (90\% confidence limits).
\label{fig7} }

\figcaption{A comparison of the PN spectra
of EZ CMa and WR110. Spectra are rebinned to a minimum of
20 counts per bin.
\label{fig8} }


\clearpage

\begin{figure}
\figurenum{1}
\epsscale{0.6}
\plotone{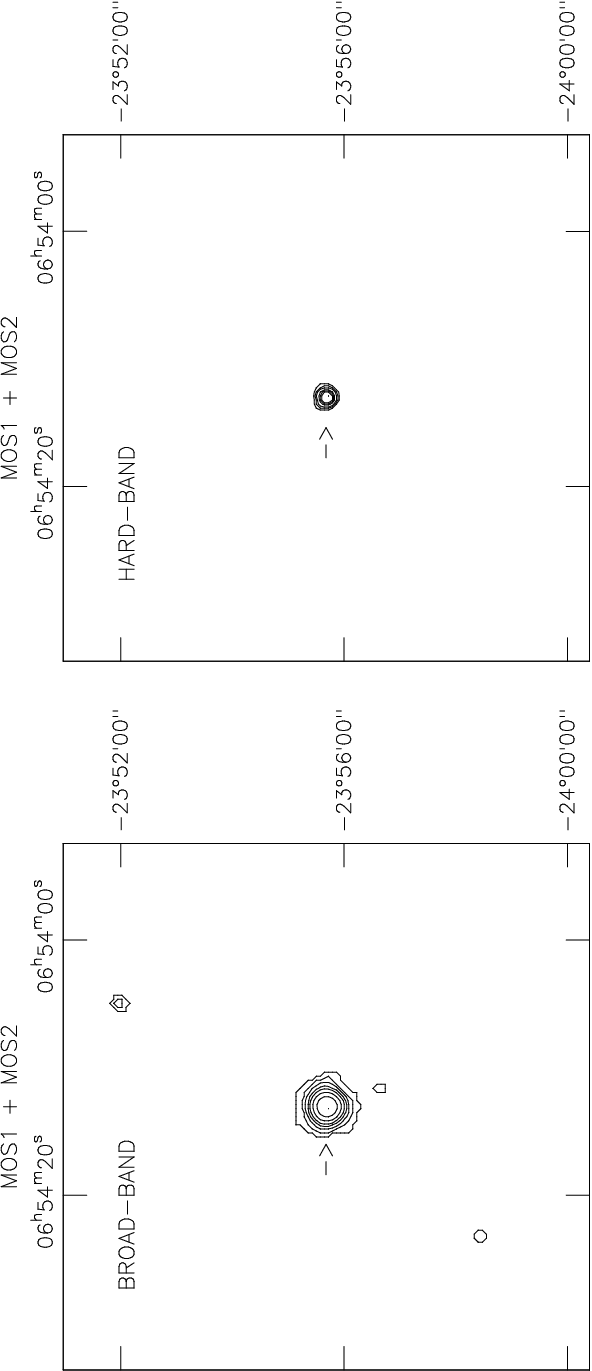}
\caption{}
\end{figure}
 
\clearpage

\begin{figure}
\figurenum{2}
\epsscale{1.0}
\plotone{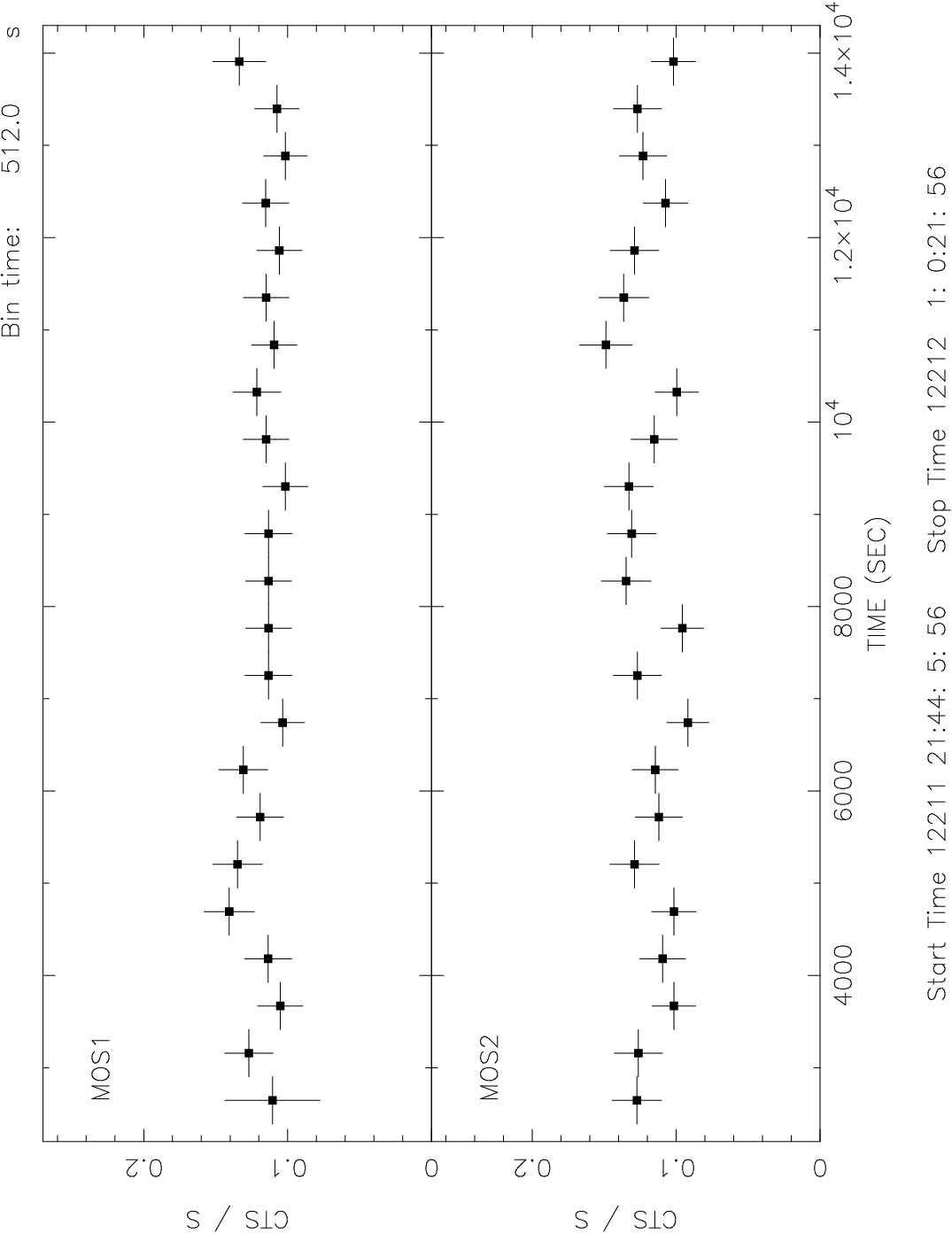}
\caption{}
\end{figure}

\clearpage

\begin{figure}
\figurenum{3}
\epsscale{0.9}
\plotone{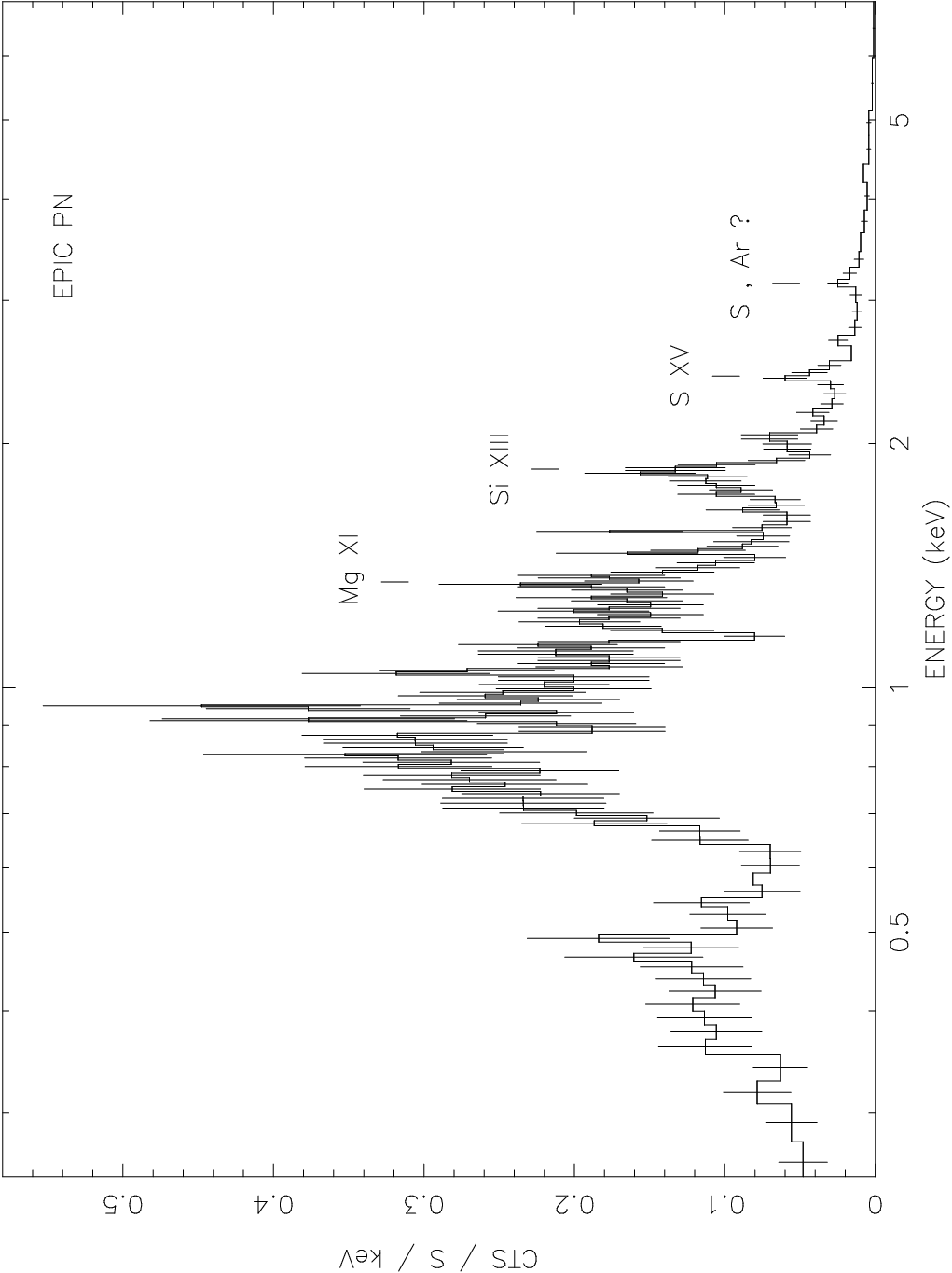}
\caption{}
\end{figure}

\clearpage

\begin{figure}
\figurenum{4}
\epsscale{0.9}
\plotone{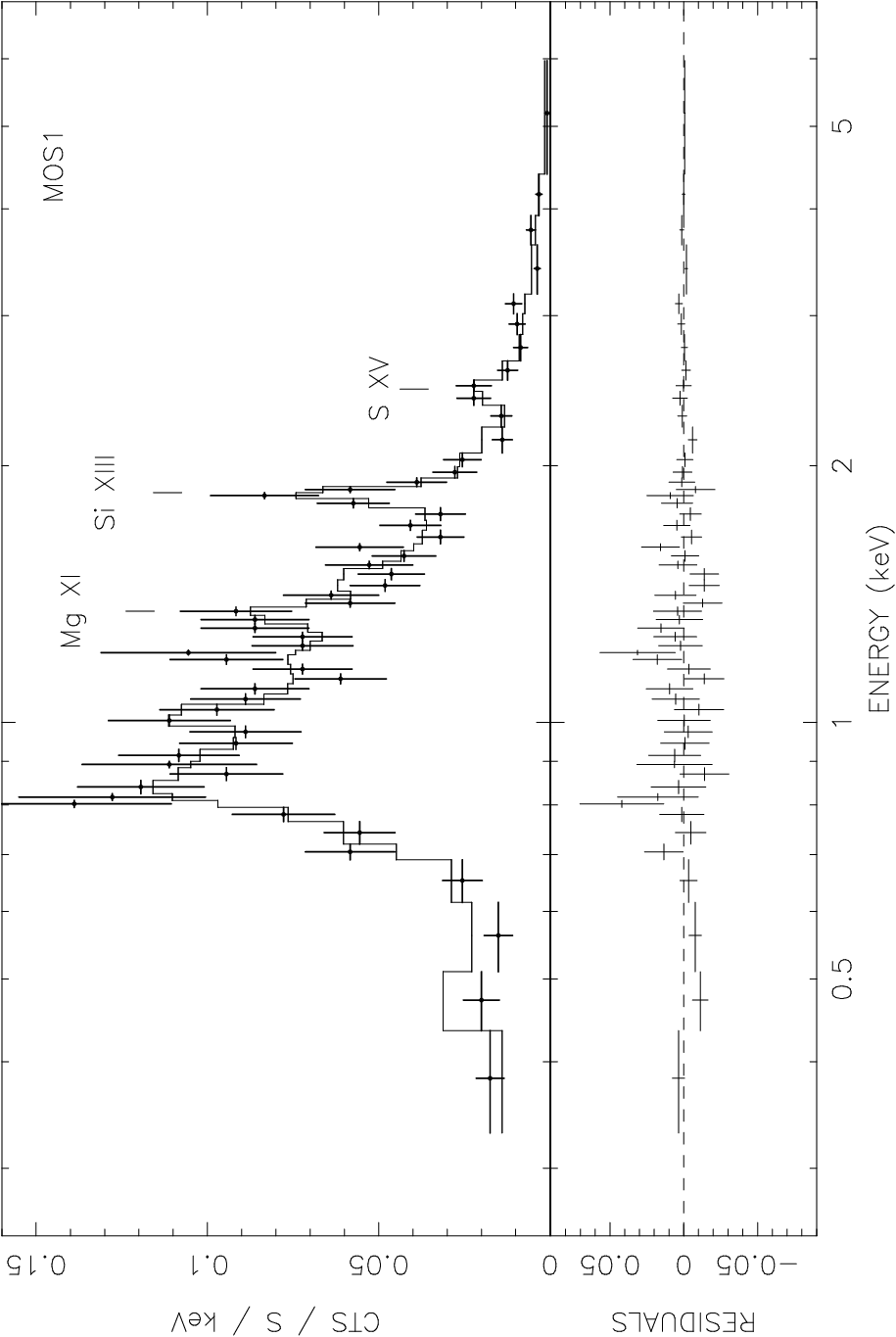}
\caption{}
\end{figure}

\clearpage

\begin{figure}
\figurenum{5}
\epsscale{0.9}
\plotone{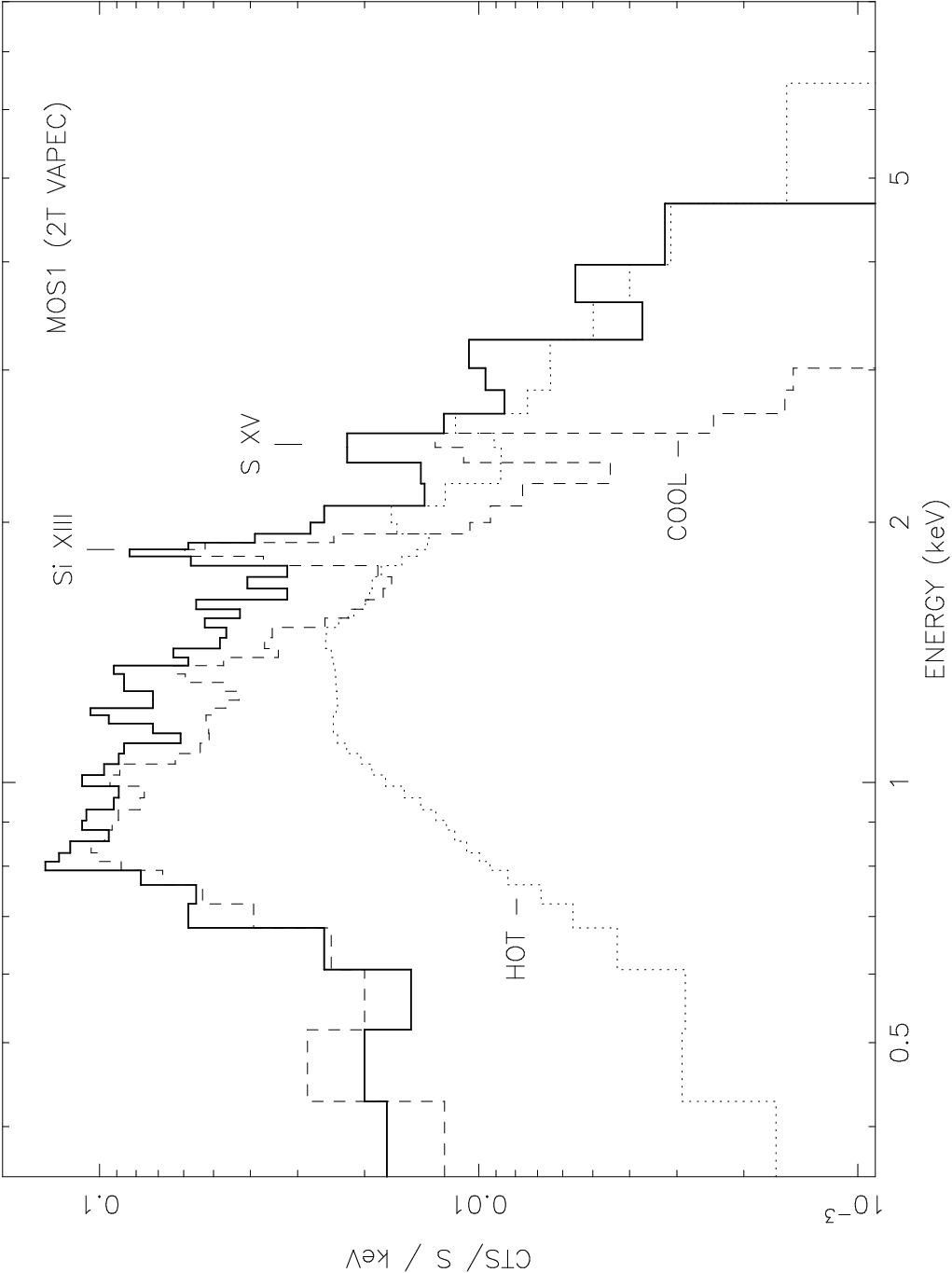}
\caption{}
\end{figure}

\clearpage

\begin{figure}
\figurenum{6}
\epsscale{1.0}
\plotone{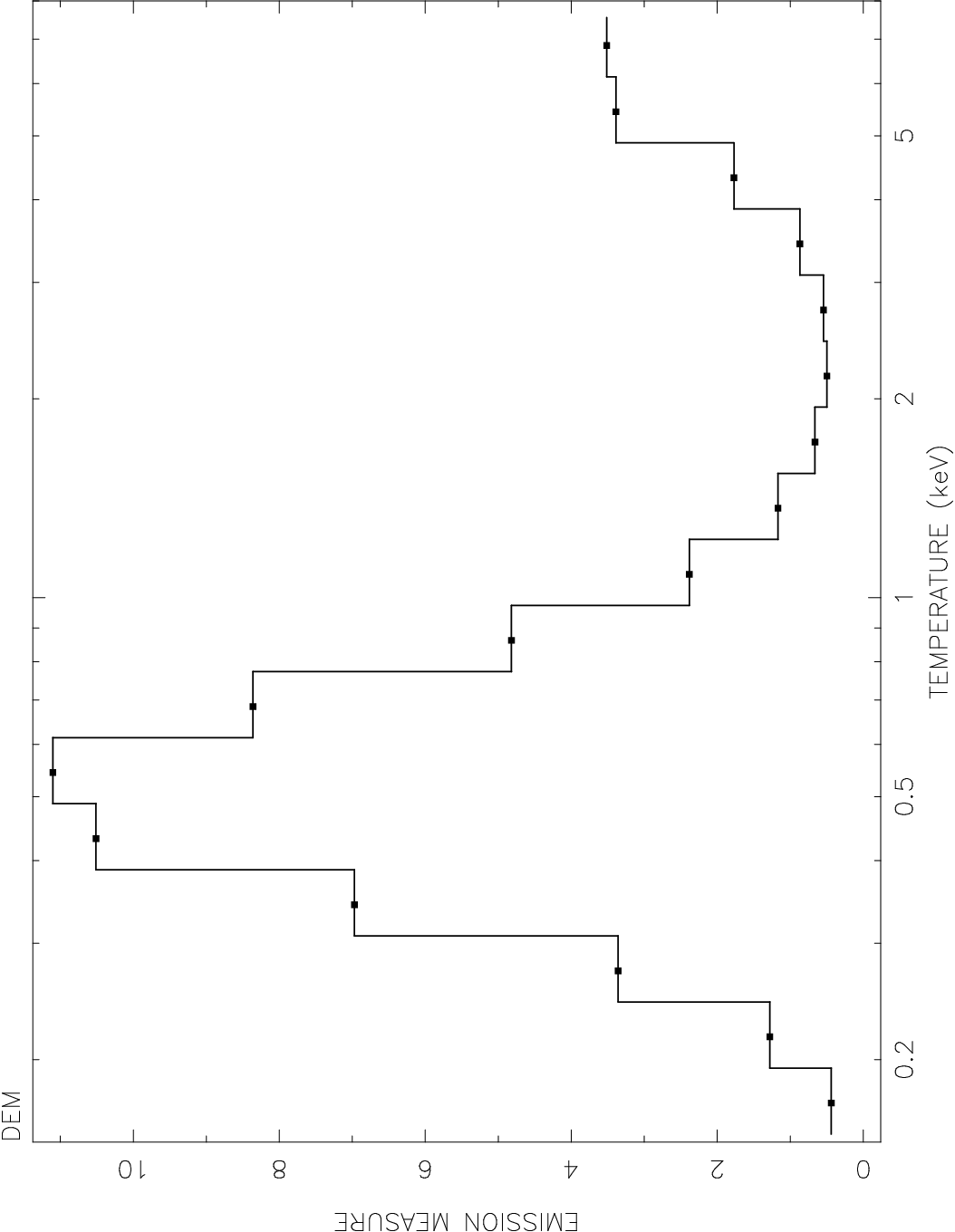}
\caption{}
\end{figure}

\clearpage

\begin{figure}
\figurenum{7}
\epsscale{0.9}
\plotone{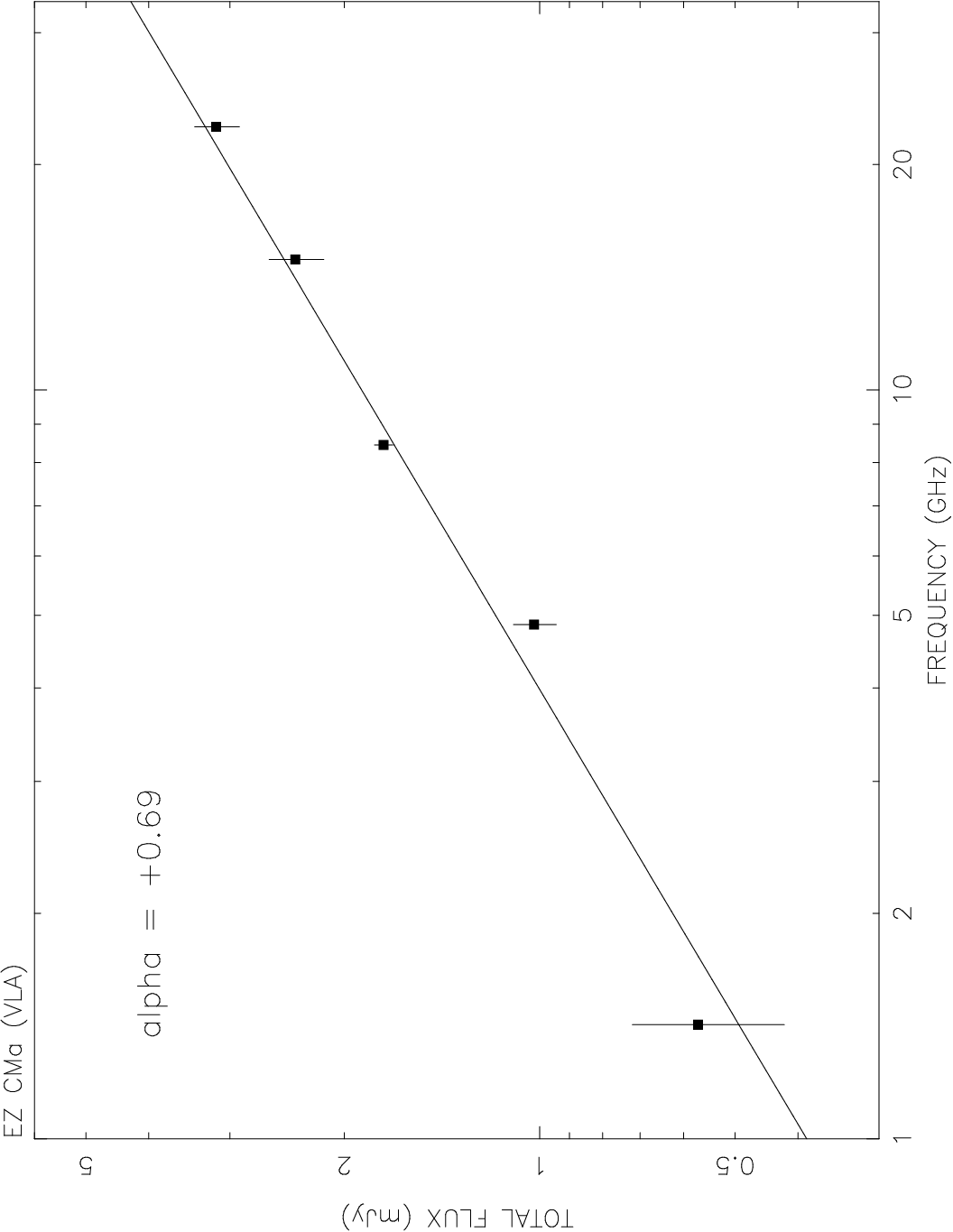}
\caption{}
\end{figure}

\clearpage

\begin{figure}
\figurenum{8}
\epsscale{0.9}
\plotone{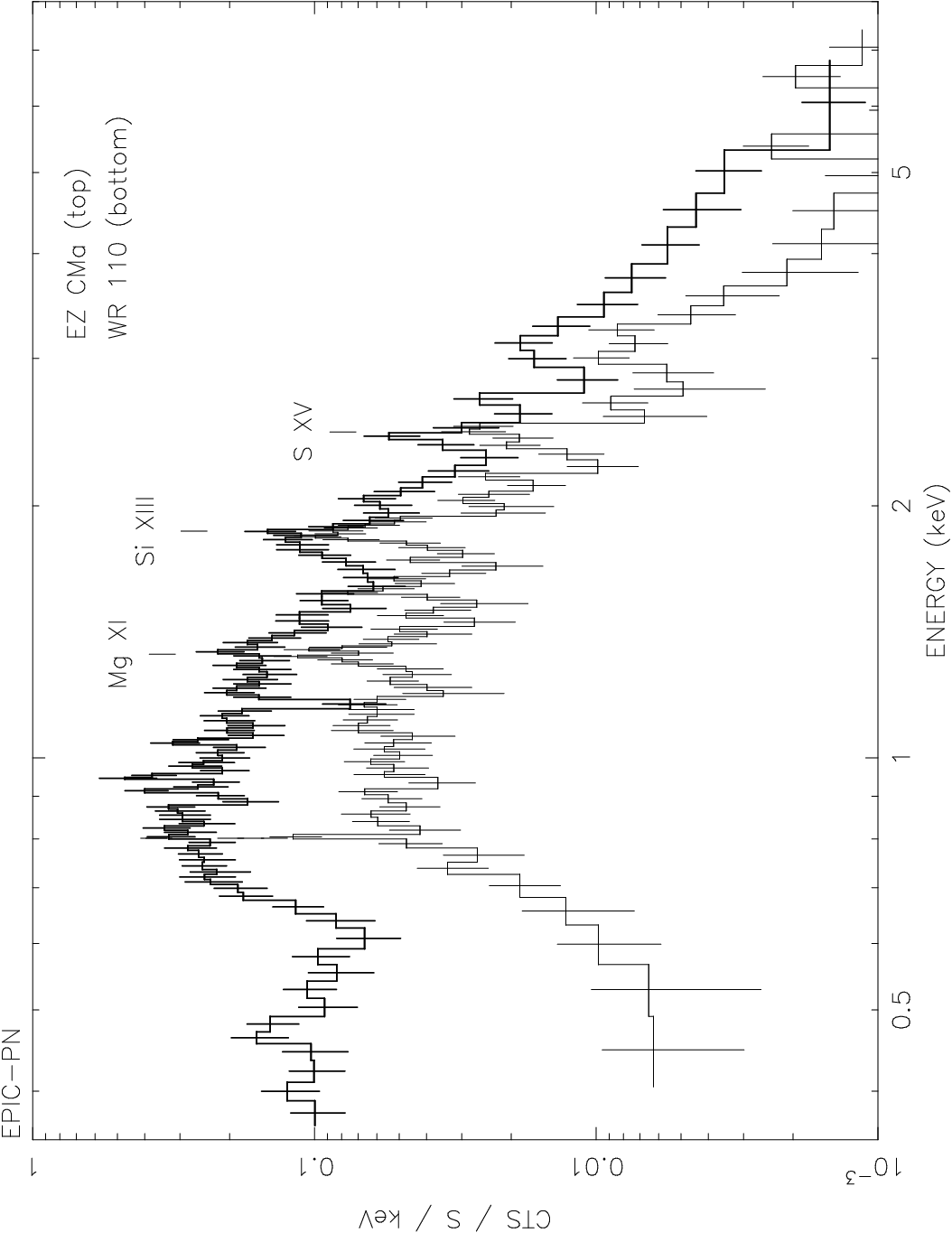}
\caption{}
\end{figure}


\begin{thebibliography}{}
\bibitem[]{} Abbott, D.C., Bieging, J.H., Churchwell, E., 
\& Torres, A.V. 1986, \apj, 303, 239
\bibitem[]{} Altenhoff, W.J., Thum, C., \& Wendker, H.J., 1994,
\aap, 281, 161
\bibitem[]{} Anders, E., \& Grevesse, N. 1989, \gca, 53, 197 (AG89)
\bibitem[]{} Arnaud, K.A. 1996, in Astronomical Data Analysis 
Software and Systems V, eds. G. Jacoby \&  J. Barnes, 
(San Francisco: ASP), 101, 17 
\bibitem[]{} Baum, E., Hamann, W.-R., Koesterke, L. \&
Wessolowski, U. 1992, \aap, 266, 402
\bibitem[]{} Cassinelli, J.P., Miller, N.A., Waldron, W.A., 
MacFarlane, J.J., \& Cohen, D.H., 2001, \apj, 554, L55
\bibitem[]{} Cherepashchuk, A.M. 1976, Soviet Astr. Letters, 2, 138
\bibitem[]{} Cohen, D. et al. 2002, \apj, in prep.
\bibitem[]{} Contreras, M.E. \& Rodriguez, L.F. 2000, Rev. Mex.
de Astron. y Astrof., 36, 135 
\bibitem[]{} Davidson, K. \& Ostriker, J.P., 1973, \apj, 179, 585
\bibitem[]{} Duijsens, M.F.J., van der Hucht, K.A., van Genderen,
A.M., Schwarz, H.E., Linders, H.P.J., \& Kolkman, O.M. 1996,
\aaps, 119, 37
\bibitem[]{} Dumm, T. et al. 2002, to appear in New Visions of the
X-ray Universe in the XMM-Newton and Chandra Era, ed. F. Jansen,
ESA SP-488, in press
\bibitem[]{} Feldmeier, A., Kudritzki. R.-P., Palsa, R.
Pauldrach, A.W.A., \& Puls, J., 1997, \aap, 320, 899
\bibitem[]{} Firmani, C., Koenigsberger, G., Bisiacchi, G.F.,
Moffat, A.F.J., \& Isserstedt, J. 1980, \apj, 239, 607
\bibitem[]{} Gayley, K.G. \& Owocki, S.P., 1995, \apj, 446, 801
\bibitem[]{} Gayley, K.G., Owocki, S.P., \& Cranmer, S.R., 1997,
\apj, 475, 786
\bibitem[]{} Georgiev, L.N., Koenigsberger, G., Ivanov, M.M.,
St.-Louis, N., \& Cardona, O. 1999, \aap, 347, 583
\bibitem[]{} Gorenstein, P. 1975, \apj, 198, 95
\bibitem[]{} Hamann, W.-R. \& Koesterke, L. 1998, \aap, 333, 251
\bibitem[]{} Harries, T.J., Howarth, I.D., Schulte-Ladbeck, R.E., 
\& Hillier, D.J. 1999, \mnras, 302, 499
\bibitem[]{} Hillier, D.J., 1987, \apjs, 63, 965 
\bibitem[]{} Hillier, D.J., Kudritzki, R.P., Pauldrach, A.W., 
Baade, D., Cassinelli, J.P., Puls, J., \& Schmitt, J.H.M.M., 1993,
\aap, 276, 117
\bibitem[]{} Hogg, D.E. 1989, \aj, 98, 282
\bibitem[]{} Howarth, I.D. \& Schmutz, W. 1995, \aap, 294, 529
\bibitem[]{} Ignace, R., Oskinova, L.M., \& Foullon, C., 2000,
             \mnras, 318, 214
\bibitem[]{} Jansen, F. et al., 2001, \aap, 365, L1
\bibitem[]{} Kahn, S.M. et al., 2001, \aap, 365, L312
\bibitem[]{} Lamontagne, R., Moffat, A.F.J., \& Lamarre, A. 1986,
\aj, 91, 925
\bibitem[]{} Leitherer, C. \& Robert, C. 1991, \apj, 377, 629
\bibitem[]{} Lemen, J.R., Mewe, R., Schrijver, C.J., \& Fludra, A., 1989,
\apj, 341, 484
\bibitem[]{} Lucy, L.B., 1982, \apj, 255, 286
\bibitem[]{} Lucy, L.B. \& White, R.L., 1980, \apj, 241, 300
\bibitem[]{} Luo, D., McCray, R., \& MacLow, M-M., 1990, \apj, 362, 267
\bibitem[]{} MacFarlane, J.J. et al. 1991, ApJ, 380, 564
\bibitem[]{} Moffat, A.F.J., Firmani, C., McLean, I.S., \&
Seggewiss, W. 1982, in Wolf-Rayet Stars: Observations, Physics,
Evolution, eds. C. de Loore \& A. Willis (Dordrecht: Reidel), 577
\bibitem[]{} Morrison, R. \& McCammon, D. 1983, \apj, 270, 119
\bibitem[]{} Owocki, S.P., Castor, J.I., \& Rybicki, G.B., 1988,
\apj, 335, 914
\bibitem[]{} Owocki, S.P. \& Cohen, D.H. 2001, \apj, 559, 1108
\bibitem[]{} Perryman, M.A.C. et al. 1997, \aap, 323, 49
\bibitem[]{} Pozzo, M., Jeffries, R.D., Naylor, T.,
Totten, E.J., Harmer, S., \& Kenyon, M. 2000, 
\mnras, 313, L23
\bibitem[]{} Prilutskii, O., Usov, V.V., 1976, SvA-AJ, 20,2
\bibitem[]{} Prinja, R.K., Barlow, M.J., \& Howarth, I.D., 
1990, \apj, 361, 607
\bibitem[]{} Runacres, M.C. \& Owocki, S.P., 2002, \aap, 381, 1015
\bibitem[]{} Schmutz, W. 1997, \aap, 321, 268
\bibitem[]{} Skinner, S.L., G\"{u}del, M., Schmutz, W., \& Stevens, I.R.,
2001, \apj, 558, L113
\bibitem[]{} Skinner, S.L., Itoh, M., \& Nagase, F. 1997,
New Astron., 3, 37
\bibitem[]{} Skinner, S.L., Zhekov, S.,  G\"{u}del, M., \& Schmutz, W.
2002, \apj, 572, in press (SZGS02)
\bibitem[]{} Stevens, I.R. \& Willis, A.J. 1988, \mnras, 234, 783
\bibitem[]{} St.-Louis, N., Dalton, M.J., Marchenko, S.V.,
Moffat, A.F.J., \& Willis, A.J., 1995, \apj, 452, L57
\bibitem[]{} Str\"{u}der, L. et al., 2001, \aap, 365, L18
\bibitem[]{} Turner, M.J.L. et al., 2001, \aap, 365, L27
\bibitem[]{} Usov, V.V. 1992, \apj, 389, 635
\bibitem[]{} van der Hucht, K.A., 2001, New Ast. Rev., 45, 135 (vdH01)
\bibitem[]{} van der Hucht, K.A., Cassinelli, J.P., \& Williams, P.M., 
1986, \aap, 168, 111 (VCW86)
\bibitem[]{} Waldron, W. \& Cassinelli, J.P., 2001, \apj, 548, L45
\bibitem[]{} White, R.L. \& Long, K.S., 1986, \apj, 310, 832
\bibitem[]{} Willis, A.J. 1996, \apss, 237, 145
\bibitem[]{} Willis, A.J. \& Stevens, I.R., 1996, \aap, 310, 577 (WS96)
\bibitem[]{} Wright, A.E. \& Barlow, M.J., 1975, \mnras, 170, 41
\end{thebibliography}
\end{document}